\documentclass[11pt]{article}
\usepackage{moriond,epsfig}

\def\Journal#1#2#3#4{{#1} {\bf #2}, #3 (#4)}

\def\ApJ{{\em Astrophys. Journal}}
\def\CQG{{\em Class. Quant. Grav.}}
\def\AJ{{\em Astronom. Journal}}
\def\AA{{\em Astron. \& Astrophys.}}
\def\MNRAS{{\em Mont. Not. Roy. Astron. Soc.}}

\def\be{\begin{equation}}
\def\ee{\end{equation}}
\def\bea{\begin{eqnarray}}
\def\eea{\end{eqnarray}}

\begin{document}
\vspace*{4cm}
\title{SNIa DATA TO PROBE THE COSMOLOGICAL PRINCIPLE}

\author{ M. N. C\'EL\'ERIER }

\address{D\'epartement d'Astrophysique Relativiste et de Cosmologie, \\
Observatoire de Paris-Meudon, 5 place Jules Janssen, 92195 
Meudon C\'edex, France}

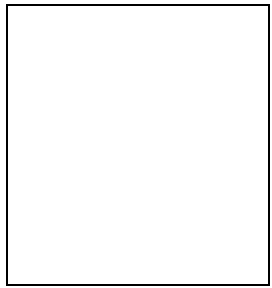
\maketitle\abstracts{Recently collected SNIa data have been used to 
address the problem of measuring the cosmological 
parameters of the universe. Analysed in the 
framework of homogeneous models, they have 
yielded, as a primary result, a strictly 
positive cosmological constant. However, a 
straight reading of the published measurements, 
conducted with no a priori idea of which model 
would best describe our universe, at least 
up to redshifts z=1, does not exclude the 
possibility of ruling out the Cosmological 
Principle  - and cosmological constant - 
hypotheses. It is here shown how the large scale 
(in)homogeneity of this part of the universe can be 
tested on our past light cone, using the 
magnitude-redshift relation, provided sufficiently
accurate data from sources at redshifts 
approaching z=1 would be available. }

\section{Introduction}

The widespread belief in large scale spatial homogeneity 
for our universe proceeds from an hypothesis brought to the status of 
Cosmological Principle. Its justification is based on two arguments: 
1.The isotropy (or quasi-isotropy) of the CMBR around us. 
2.The Copernican assumption that, as our location must not be special, 
this observed isotropy must be the same from any other point of the 
universe. However, of the two above arguments, only the first 
is observation grounded. The second, purely philosophical, cannot be 
directly verified. \\

Another common belief is that the inhomogeneities observed in the 
universe can be consistently smoothed out in an averaged homogeneous 
model. {\em This is absolutly wrong} and usualy leads to 
improper uses of FLRW relations \cite{ksl,mbhe}. \\

The discovery of high-redshift SNIa and their use 
as standard candles have resurected interest in the 
magnitude-redshift (M-R) relation as a tool to measure the 
cosmological parameters of the universe. 
Data recently collected by two survey teams (the Supernova 
Cosmology Project and the High-z Supernova Search Team), and analysed 
in the framework of homogeneous FLRW cosmological models, have 
yielded, as a primary result, a strictly positive cosmological 
constant, of order unity \cite{ri,pe}. 
If these results were to be confirmed, it would be 
necessary to explain how $\Lambda$ is so small, yet non zero. Hence, 
a revolutionary impact in both cosmology and particle physics.\\

The purpose is here, 
assuming every source of potential bias or systematic uncertainties 
has been correctly taken into account in the data collecting, to 
probe the large scale homogeneity on our past 
light cone available with the SNIa measurements, thus testing the 
Cosmological Principle and cosmological constant hypotheses.

\section{The magnitude-redshift relation to probe homogeneity}

Consider any cosmological model for which the 
luminosity distance $D_L$ is a function of the redshift $z$ and of the 
parameters $cp$ of the model. Assume that $D_L$ is Taylor expandable 
near the observer, i.e. around $z=0$,

\begin{equation}
D_L(z;cp) = \left(dD_L\over dz\right)_{z = 0}z \, + \, {1\over 2}
\left(d^2D_L\over dz^2\right)_{z = 0}z^2
+ \, {1\over 6}\left(d^3D_L \over dz^3\right)_{z = 0}z^3 \, + \, 
{1\over 24}\left(d^4D_L\over dz^4\right)_{z = 0}z^4 + {\cal O}(z^5) , 
\label{eq:dl}
\end{equation}

Therefore, the apparent bolometric magnitude $m$ of a standard candle of 
absolute bolometric magnitude $M$ is also a function of $z$ and $cp$. In 
megaparsecs,

\begin{equation}
m=M+5logD_L(z;cp)+25
\label{eq:mr}
\end{equation}

Luminosity-distance measurements of such sources at increasing 
redshifts $z<1$ thus yield values for the coefficients at 
increasing order in the above expansion. For cosmological models with 
high, or infinite, number of free parameters, the observations only 
produce constraints upon the parameter values near the observer. 
For cosmological models with few constant parameters, giving 
independent contributions to each coefficient in the expansion, 
the observed M-R relation provides a way to test the validity of the 
model, and, if valid, to evaluate its parameters. \\

For Friedmann models precisely, the expansion coefficients $D_L^{(i)}$ 
are independent functions of the three constant parameters $H_0$, 
$\Omega_M$ and $\Omega_\Lambda$, and can be derived from the 
expression of $D_L$ \cite{ce}. 
Therefore, accurate luminosity-distance measurements 
of three samples of same order redshift SNIa would yield 
values for $D_L^{(1)}$, $D_L^{(2)}$ and $D_L^{(3)}$ and 
thus select a triplet of numbers for the model parameters $H_0$, 
$\Omega_M$ and $\Omega_\Lambda$ \footnote{In fact, $H_0$ 
can be hidden in the magnitude zero-point ${\cal M}\equiv M-5 
\log H_0+25$. The SCP team calibrates ${\cal M}$ from the data of a 
low-redshift sample and claims direct fitting of $\Omega_M$ and 
$\Omega_\Lambda$. The HzSST calibrates $M$ and claims, as a bonus, an 
estimate of $H_0$.}. \\

1. If $\Omega_M<0$, which leads to physical inconsistency - 
but cannot be excluded from the current data (see e.g. Fig. 6 of 
Riess {\em et al.} \cite{ri}, where the permitted ellipses can be 
extended to the $\Omega_M<0$ region) - the homogeneity assumption 
have to be ruled out at this stage. \\

2. If $\Omega_M>0$, the triplet can be used to provide 
a prediction for the value of the fourth order 
coefficient $D_L^{(4)}$. Now, if further observations at redshifts 
approaching unity are made, $D_L^{(4)}$ can be 
determined and compared to its predicted value, thus providing 
a test of the FLRW model. \\

If the ongoing surveys were to discover more distant sources, at 
redshifts higher than unity, the Taylor expansion would no longer be 
valid. Therefore, we should consider the fit of the Hubble diagram for 
accurately measured sources at every available scale of redshift
\cite{ce}. 

\section{Simplified inhomogeneous models}

Two statements remain to be proved at this stage: 
1. The ruling out of the Cosmological Principle is 
not a purely academical possibility. Physically robust 
inhomogeneous models exist which can verify any observed 
M-R relation. 
2. A non-zero cosmological constant is not mandatory, as 
$\Lambda=0$ inhomogeneous models can mimic 
$\Lambda\neq 0$ Friedmann ones.

\subsection{Example: LTB models with $\Lambda = 0$}

Lema\^itre \cite{le}-Tolman \cite{to}-Bondi \cite{bo} (LTB) models 
are spatially 
spherically symmetrical solutions of Einstein's equations with dust 
as the source of gravitationnal energy. They can thus be retained to 
roughly represent a quasi-isotropic 
universe in the matter dominated area. Indeed, a spherically symmetric 
model may be regarded as describing data that have been averaged 
over the whole sky, but not over distance \cite{mbhe}. \\

The line-element, in comoving coordinates ($r,\theta,\varphi$) and
proper time $t$, is (with $c = 1$)
\begin{equation}
ds^2 = -dt^2 + S^2(r,t)dr^2 + R^2(r,t)(d\theta^2 + \sin^2 \theta 
d\varphi^2) . \label{eq:le}
\end{equation}

Einstein's equations with $\Lambda = 0$ imply that the metric 
coefficients are functions 
of the time-like $t$ and radial $r$ coordinates, and of two 
independent functions of $r$, which play the role of the model 
parameters ($cp$). 
The radial luminosity distance is
\begin{equation}
D_L = (1+z)^2R . \label{eq:rdl}
\end{equation}

The $D_L$ expansion coefficients follow, as independent functions 
of the derivatives of the model parameters, evaluated at the  
observer ($z = 0$). These parameters, which are implicit 
functions of $z$, through the null geodesic equations, are present 
in each coefficient $D_L^{(i)}$ with derivatives up to the  $i$th order 
\cite{ce}. LTB models are thus completly degenerate with respect to 
any M-R relation. \\

The LTB example has been chosen for calculation simplicity. The complete 
degeneracy obtained by relaxing only one 
symmetry of the corresponding homogeneous model, namely spatial 
translation, allows to infer that a less symmetric inhomogeneous model, 
where the spherical symmetry would for instance also be relaxed, would 
provide an even higher degree of degeneracy.

\subsection{Illustration: the flat ($\Lambda=0$) LTB model}

Flat ($\Lambda=0$) LTB solutions can be characterized by only one 
arbitrary function of $r$, namely $t_0(r)$,  usually interpreted, for 
cosmological use, as a Big-Bang singularity surface. The $D_L^{(i)}$ 
coefficients can thus be expressed in term of the successive derivatives 
of $t_0(r)$, up to the $i$th order. \\

A comparison with the corresponding FLRW coefficients gives the 
following relations:
\begin{eqnarray}
\Omega _M \longleftrightarrow 1+5\, {t'_0(0)\over \left(9GM_0\over 
2\right)^{1\over 3}t_p^{2\over 3}}+{29\over 4}\, 
{t'^2_0(0)\over \left(9GM_0\over 2\right)^{2\over 3}t_p^{4\over 3}}
+{5\over 2}\, {t''_0(0)\over \left(9GM_0\over 2\right)^{2\over 3}
t_p^{1\over 3}} , \label{eq:om} \\
\Omega _\Lambda \longleftrightarrow -\, {1\over 2}\, {t'_0(0)\over 
\left(9GM_0\over 2\right)^{1\over 3}t_p^{2\over 3}}+{29\over 8}\, 
{t'^2_0(0)\over \left(9GM_0\over 2\right)^{2\over 3}
t_p^{4\over 3}}
+{5\over 4}\, {t''_0(0)\over \left(9GM_0\over 
2\right)^{2\over 3}t_p^{1\over 3}} . \label{eq:ol} \\
\nonumber
\end{eqnarray}

This implies that a positive $\Lambda$ in a FLRW interpretation of the 
data at $z<1$ corresponds to a mere constraint on the model 
parameter in a flat LTB ($\Lambda =0$) interpretation. \\

For their latest published results, the SCP team  propose a FLRW 
interpretation, which they write as \cite{pe} $0.8\, \Omega _M-0.6\, 
\Omega _\Lambda \approx - \, 0.2\, \pm \, 0.1$. 
In a flat $\Lambda=0$ LTB interpretation, this becomes \footnote{ $t_p$ 
and $M_0$ are not free parameters of the model. $t_p$ is the time-like 
coordinate at the observer, and its value proceeds from the measured 
temperature at 2.73 K \cite{sc}. $M_0$ is a constant setting the scale 
of the comoving radial coordinate $r$.}

\begin{equation}
4.3\, {t'_0(0)\over \left(9GM_0\over 2\right)^{1\over 3}t_p^{2\over 3}}
+3.625\, {t'^2_0(0)\over \left(9GM_0\over 2\right)^{2\over 3}t_p^{4\over 
3}}+1.25\, {t''_0(0)\over \left(9GM_0\over 2\right)^{2\over 3}
t_p^{1\over 3}} \approx -1\, \pm \, 0.1 . \label{eq:ire}
\end{equation}

Note that, even with $\Lambda=0$, degeneracy still 
remains, as an infinite number of $\{t_0'(0), t_0''(0)\}$ couples can 
verify Eq.(\ref{eq:ire}).

\section{Conclusions}

Provided SNIa would be confirmed as good standard candles, 
data from this kind of sources at redshifts approaching unity could, 
in a near future, be used to test the homogeneity assumption 
on our past light cone. 
Would this assumption be discarded by the shape of the measured 
M-R relation, inhomogeneous solutions could provide 
good alternative models, as they are completly degenerate with respect 
to any of these relations, even with a vanishing cosmological 
constant. 
Would a FLRW type distance-redshift relation be observed, it 
would not be enough to strongly support the Cosmological Principle, as 
the possibility for an inhomogeneous model to mimic such a relation 
could not be excluded. \\
 
Therefore, at the current stage reached by the observations, a 
non-zero $\Lambda$ is not mandatory, as $\Lambda=0$ inhomogeneous 
models can mimic a $\Lambda\neq 0$ FLRW M-R relation. \\

In any case, to consolidate the robustness of future 
M-R tests, it would be worth confronting their results 
with the full range of available cosmological data, {\em analysed in 
a model independent way}.


\section*{References}
\begin{thebibliography}{99}
\bibitem{ksl}H. Kurki-Suonio and E. Liang, 
\Journal{\ApJ}{390}{5}{1992}.

\bibitem{mbhe}N. Mustapha, B.A.C.C. Bassett, C. Hellaby 
and G.F.R. Ellis, \Journal{\CQG}{15}{2363}{1998}.

\bibitem{ri}A.G. Riess {\it et al}, 
\Journal{\AJ}{116}{1009}{1998}.

\bibitem{pe}S. Perlmutter {\it et al}, 
\Journal{\ApJ}{517}{565}{1999}.

\bibitem{ce}M.N. C\'el\'erier, \Journal{\AA}{353}{63}{2000}.

\bibitem{le}G. Lema\^itre, \Journal{\em Ann. Soc. Sci. Bruxelles}
{A53}{51}{1933}.

\bibitem{to}R.C. Tolman, 
\Journal{\em Proc. Nat. Acad. Sci.}{20}{169}{1934}.

\bibitem{bo}H. Bondi, \Journal{\MNRAS}{107}{410}{1947}.

\bibitem{sc}J. Schneider and M.N. C\'el\'erier, 
\Journal{\AA}{348}{25}{1999}.



\end{thebibliography}
\end{document}